\documentclass[11pt]{article}
\usepackage{amsmath,amssymb,amsthm,amsxtra,overpic,bbm,bm,epsfig,ulem,color,multirow}
\usepackage{braket}

\begin{document}

\begin{center}
{\Large  Neutrino mass anarchy and leptogenesis  }
\end{center}

\vspace{0.05cm}

\begin{center}
{\bf Qi An, \bf Yan Shao\footnote{Corresponding author: shaoyan0998@163.com}, \bf Zhen-hua Zhao\footnote{Co-corresponding author: zhaozhenhua@lnnu.edu.cn}} \\
{ $^1$ Department of Physics, Liaoning Normal University, Dalian 116029, China \\
$^2$ Center for Theoretical and Experimental High Energy Physics, \\ Liaoning Normal University, Dalian 116029, China }
\end{center}

\vspace{0.2cm}

\begin{abstract}
In this paper, we investigate leptogenesis under the neutrino mass anarchy hypothesis in both type-I and type-II seesaw models. We first revisit the corresponding study in the type-I seesaw framework with two improvements: in contrast to Ref.~\cite{Lu:2014cla}, where an approximate $U(1)$ flavor symmetry was imposed to ensure sizable hierarchies among the right-handed neutrino masses and Yukawa couplings, we adopt a fully general anarchy scenario with completely random and structureless neutrino mass and Yukawa matrices; moreover, given the crucial role of lepton flavors in both the generation and washout of the lepton asymmetry, flavor effects are consistently incorporated throughout our analysis. We then extend our investigation to the type-II seesaw framework, in which leptogenesis proceeds via the out-of-equilibrium decays of a scalar triplet.
\end{abstract}

\newpage

\section{Introduction}

The discovery of neutrino oscillations has established that neutrinos are massive and that lepton flavors are mixed \cite{xing}. This observation provides clear evidence for physics beyond the Standard Model (SM), in which neutrinos are assumed to be massless. To account for the origin of neutrino masses, various theoretical frameworks beyond the SM have been proposed. Among the various mechanisms proposed to account for neutrino masses, the seesaw mechanism offers a particularly attractive explanation by introducing new heavy states whose exchange generates small Majorana masses for the light neutrinos.

The simplest realization is the type-I seesaw mechanism, in which heavy right-handed neutrinos $N^{}_I$ (with $I=1, 2, ...$) are added to the SM \cite{seesaw1}--\cite{seesaw5}. In this model, the Yukawa interactions between the right-handed and left-handed neutrinos generate the Dirac neutrino mass matrix $M^{}_{\rm D}= Y^{}_{\nu} v$, where $Y^{}_{\nu}$ denotes the neutrino Yukawa coupling matrix and $v = 174$ GeV is the vacuum expectation value of the Higgs field. In addition, the right-handed neutrinos possess the Majorana mass matrix $M^{}_{\rm R}$. Under the seesaw condition $M^{}_{\rm R} \gg M^{}_{\rm D}$, the effective mass matrix for light neutrinos is obtained as
\begin{eqnarray}
M^{}_{\nu} \simeq - M^{}_{\rm D} M^{-1}_{\rm R} M^{T}_{\rm D} \;.
\label{1}
\end{eqnarray}
Diagonalization of Eq.~(\ref{1}) by means of the PMNS mixing matrix $U$ \cite{pmns1, pmns2} gives the light neutrino masses:
\begin{eqnarray}
U^\dagger M^{}_{\nu} U^* = D^{}_\nu= {\rm diag}(m^{}_1, m^{}_2, m^{}_3) \;,
\label{2}
\end{eqnarray}
with $m^{}_i$ (for $i=1, 2, 3$) being three light neutrino masses. In the standard parametrization, $U$ depends on three mixing angles $\theta^{}_{ij}$ (for $ij=12, 13, 23$), one Dirac CP phase $\delta$ and two Majorana CP phases $\rho$ and $\sigma$ as
\begin{eqnarray}
U  = \left( \begin{matrix}
c^{}_{12} c^{}_{13} & s^{}_{12} c^{}_{13} & s^{}_{13} e^{-{\rm i} \delta} \cr
-s^{}_{12} c^{}_{23} - c^{}_{12} s^{}_{23} s^{}_{13} e^{{\rm i} \delta}
& c^{}_{12} c^{}_{23} - s^{}_{12} s^{}_{23} s^{}_{13} e^{{\rm i} \delta}  & s^{}_{23} c^{}_{13} \cr
s^{}_{12} s^{}_{23} - c^{}_{12} c^{}_{23} s^{}_{13} e^{{\rm i} \delta}
& -c^{}_{12} s^{}_{23} - s^{}_{12} c^{}_{23} s^{}_{13} e^{{\rm i} \delta} & c^{}_{23}c^{}_{13}
\end{matrix} \right) \left( \begin{matrix}
e^{{\rm i}\rho} &  & \cr
& e^{{\rm i}\sigma}  & \cr
&  & 1
\end{matrix} \right) \;,
\label{3}
\end{eqnarray}
where the abbreviations $c^{}_{ij} = \cos \theta^{}_{ij}$ and $s^{}_{ij} = \sin \theta^{}_{ij}$ have been employed. These mixing parameters (excluding the Majorana CP phases) are measured through neutrino oscillation experiments, which are also sensitive to the independent neutrino mass-squared differences: $\Delta m^2_{21}$ and $\Delta m^2_{31}$, where $\Delta m^2_{ij} \equiv m^2_i - m^2_j$ has been defined. The global analysis of neutrino oscillation date is available in Refs.~\cite{global1, global2}. However, neutrino oscillation experiments cannot measure the absolute neutrino mass scale and Majorana CP phases. The values of these parameters must be inferred from non-oscillatory experiments.

In theoretical studies, the observed pattern of neutrino mixing is often interpreted as a consequence of an underlying flavor symmetry in the lepton sector \cite{FS1}-\cite{FS4}. Such symmetries can limpose specific textures on the neutrino mass matrix, providing possible explanations for the observed mixing parameters. An alternative viewpoint is provided by the neutrino mass anarchy hypothesis \cite{Hall:1999sn}, in which the neutrino mass matrices are assumed to have no special flavor structure and are instead treated as random matrices. A crucial aspect of this framework is the choice of statistical measure: to ensure that physical predictions are independent of the flavor basis, the random matrix ensemble must satisfy basis invariance \cite{Haba:2000be}. Further assuming statistical independence of the matrix elements naturally leads to a Gaussian distribution, which has been widely used to study the statistical properties of neutrino masses and mixing parameters \cite{deGouvea:2003xe, deGouvea:2012ac}.

Besides explaining the neutrino masses, the seesaw framework also provides a natural mechanism for generating the baryon--antibaryon asymmetry of the Universe through leptogenesis \cite{leptogenesis}-\cite{Lreview4}. In this scenario, CP-violating interactions involving heavy Majorana neutrinos generate a lepton asymmetry in the early Universe, which is subsequently partially converted into a baryon asymmetry via electroweak sphaleron processes. The observed baryon asymmetry is quantified by \cite{planck}
\begin{eqnarray}
Y^{}_{\rm B} \equiv \frac{n^{}_{\rm B}-n^{}_{\rm \bar B}}{s} \simeq (8.69 \pm 0.04) \times 10^{-11}  \;,
\label{6}
\end{eqnarray}
where $n^{}_{\rm B}$ ($n^{}_{\rm \bar B}$) denotes the baryon (antibaryon) number density and $s$ the entropy density. Since leptogenesis is highly sensitive to the structure of neutrino mass matrices, many studies have been carried out within specific theoretical frameworks or under additional flavor symmetry assumptions. Notably, leptogenesis within the anarchy framework supplemented by an approximate $U(1)$ flavor symmetry has been explored in Ref.~\cite{Lu:2014cla}.
There, a $U(1)$ flavor symmetry is imposed to ensure sizable hierarchies among the right-handed neutrino masses (and Yukawa couplings), so that only the contribution from the lightest right-handed neutrino to leptogenesis needs to be considered.

In this paper, we revisit the corresponding study with two improvements:
above all, in contrast to Ref.~\cite{Lu:2014cla}, we will consider a fully general anarchy scenario without imposing any particular flavor symmetry; moreover, given the crucial role of flavor effects in the generation and washout of the lepton asymmetry, we consistently incorporate them in our analysis, unlike in Ref.~\cite{Lu:2014cla}, which works in the unflavored regime. To be specifical, we consider two distinct scenarios: a fully random setup in which both the Dirac and Majorana mass matrices are generated stochastically without any constraints, and a constrained scenario where we impose a cut on the random matrices to retain only those combinations that are consistent with the measured neutrino oscillation observables.
We then extend the analysis to the type-II seesaw framework, where neutrino masses originate from a scalar triplet and leptogenesis proceeds via its out-of-equilibrium decays.

The remaining parts of this paper are organized as follows. In the next section, we investigate leptogenesis in the type-I seesaw model with random neutrino mass matrices. In sections~3, we extend the analysis to the type-II seesaw model. Finally, the summary of our main results will be given in section~4.

\section{Leptogenesis in the type-I seesaw}

In this section, we investigate leptogenesis in the type-I seesaw framework within the neutrino mass anarchy hypothesis. Two scenarios are considered: in the first, both the Dirac mass matrix $M^{}_{\rm D}$ and the Majorana mass matrix $M^{}_{\rm R}$ are generated randomly; in the second, we impose an experimental cut on the randomly generated $M^{}_{\rm D}$ and $M^{}_{\rm R}$, only retaining those combinations that reproduce neutrino mixing angles consistent with experimental measurements.

In the conventional thermal leptogenesis scenario, the lepton asymmetry is generated through CP-violating decays of heavy right-handed neutrinos. This asymmetry originates from the difference between the decay rates of $N^{}_I \to L^{}_\alpha + H$ (for $\alpha = e, \mu, \tau$) and their CP-conjugate processes $N^{}_I \to \overline{L}^{}_\alpha + \overline{H}$ (with $L^{}_\alpha$ and $H$ being respectively the lepton and Higgs doublets). The corresponding flavor-specific CP asymmetries for the decays of $N^{}_I$ are given by
\begin{eqnarray}
&& \varepsilon^{}_{I \alpha} = \frac{1}{8\pi (M^\dagger_{\rm D}
M^{}_{\rm D})^{}_{II} v^2} \sum^{}_{J \neq I} \left\{ {\rm Im}\left[(M^*_{\rm D})^{}_{\alpha I} (M^{}_{\rm D})^{}_{\alpha J}
(M^\dagger_{\rm D} M^{}_{\rm D})^{}_{IJ}\right] {\cal F} \left( \frac{M^2_J}{M^2_I} \right) \right. \nonumber \\
&& \hspace{1.cm}
+ \left. {\rm Im}\left[(M^*_{\rm D})^{}_{\alpha I} (M^{}_{\rm D})^{}_{\alpha J} (M^\dagger_{\rm D} M^{}_{\rm D})^*_{IJ}\right] {\cal G}  \left( \frac{M^2_J}{M^2_I} \right) \right\} \; ,
\label{2.1}
\end{eqnarray}
with ${\cal F}(x) = \sqrt{x} \{(2-x)/(1-x)+ (1+x) \ln [x/(1+x)] \}$ and ${\cal G}(x) = 1/(1-x)$. The generated lepton asymmetry is partially washed out by inverse decays and lepton-number-violating scattering processes. A fraction of the surviving asymmetry is then converted into a baryon asymmetry via sphaleron processes, $Y^{}_{\rm B} = c \, Y^{}_{\rm L}$ with $c \simeq -1/3$. For a hierarchical right-handed neutrino spectrum, the dominant contribution typically originates from the lightest neutrino $N^{}_1$, since its related processes will effectively washout the lepton asymmetries generated from the heavier right-handed neutrinos.

For leptogenesis, the so-called flavor effects play a crucial role, depending on the temperature of the thermal bath, which is typically set by the mass scale of the decaying right-handed neutrinos. To be specific, three distinct flavor regimes can be identified \cite{flavor1, flavor2}.
(1) Unflavored regime: in the temperature range above $10^{12}$ GeV, the charged-lepton Yukawa $y^{}_\alpha$ interactions are out of equilibrium, and the lepton flavors are indistinguishable.
(2) Two-flavor regime: in the temperature range $10^{9}-10^{12}$ GeV, the $y^{}_\tau$-related interactions are in thermal equilibrium while those of $e$ and $\mu$ are not, the $\tau$ flavor becomes distinguishable from the other two flavors. Therefore, there exist two distinct flavors: the $\tau$ flavor and a coherent superposition of the $e$ and $\mu$ flavors.
(3) Three-flavor regime: in the temperature range below $10^{9}$ GeV, where the $y^{}_\mu$-related interactions are also in thermal equilibrium, all the flavors are distinguishable, thus each flavor should be treated separately.
For a hierarchical right-handed neutrino spectrum, generating sufficient baryon asymmetry requires the lightest right-handed neutrino mass to be above ${\cal O}(10^9)$ GeV \cite{DI}. Therefore, we focus on the two-flavor regime in our analysis, and in the calculations that follow, we use the ULYSSES Python package \cite{uly1} to obtain our results.

\subsection{Random generation of mass matrices without experimental cuts}

In this subsection, we generate the Dirac and Majorana mass matrices in a purely random manner without any experimental constraints. We do not apply any selection criteria based on neutrino oscillation data, focusing solely on the statistical properties of the random matrix ensemble.

Within the neutrino mass anarchy framework, the elements of $M^{}_{\rm D}$ and $M^{}_{\rm R}$ are regarded as independent random variables.
For the Dirac mass matrix $M^{}_{\rm D}$, there are nine independent complex entries, while the Majorana mass matrix $M^{}_{\rm R}$ is symmetric and therefore contains six independent complex parameters. A natural choice is to assume that each independent matrix element is drawn from a probability distribution. When the assumption of statistical independence is combined with the requirement of basis independence, the resulting statistical measure is naturally determined to be the Gaussian measure. Consequently, the probability distributions of the matrix elements can be expressed as \cite{Lu:2014cla}
\begin{eqnarray}
 &&d{M^{}_{\rm D}} \sim \prod\limits_{ij} e^{-\left|M^{}_{\rm D,ij}\right|^2} dM^{}_{\rm D,ij}
 = \left( \prod\limits_{ij} dM^{}_{\rm D,ij} \right) e^{-tr(M^{}_{\rm D} M^{\dagger}_{\rm D})}\ , \label{eq:mDgaussian} \\
 &&d{M^{}_{\rm R}} \sim \prod\limits_i e^{-\left|M^{}_{\rm R,ii}\right|^2} dM^{}_{\rm R,ii} \prod\limits_{i<j} e^{-2\left|M^{}_{\rm R,ij}\right|^2} dM^{}_{\rm R,ij}
 = \left( \prod\limits_{i \le j} dM^{}_{\rm R,ij} \right) e^{-tr(M^{}_{\rm R} M^{\dagger}_{\rm R})}\ .
\label{2.6}
\end{eqnarray}
These expressions correspond to the Gaussian ensemble widely used in random matrix theory.
In our numerical analysis, we parameterize $M^{}_{\rm D}$ ($M^{}_{\rm R}$) as the product of an overall mass scale $D^{}_0$ ($M^{}_0$) and a matrix with dimensionless entries.
In this setup, the dimensionless components of $M^{}_{\rm D}$ and $M^{}_{\rm R}$ are generated according to the aforementioned Gaussian distributions, while their overall scales $D^{}_0$ and $M^{}_0$  are treated as free parameters.

\begin{figure*}
\centering
\includegraphics[width=6.5in]{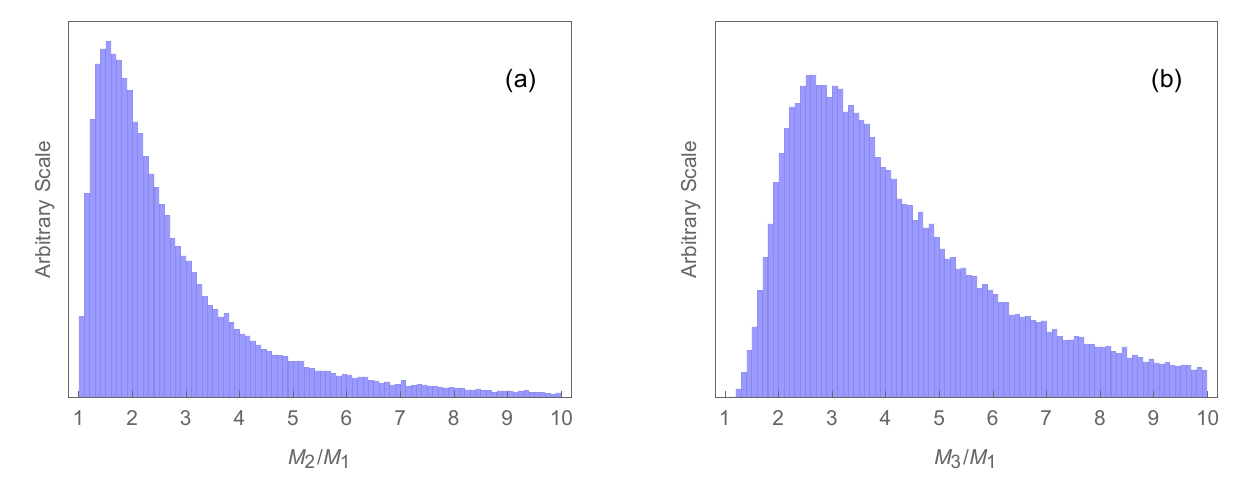}
\caption{For randomly generated right-handed neutrino mass matrix $M^{}_{\rm R}$, the probability distribution of the mass ratios $M^{}_{2}/M^{}_{1}$ and $M^{}_{3}/M^{}_{1}$.}
\label{fig1}
\end{figure*}

Based on the above formula, we first randomly generate the dimensionless components of $M^{}_{\rm R}$. Its diagonalization yields the three right-handed neutrino masses (sorted in ascending order) up to the overall scale $M^{}_0$. Repeating this procedure a sufficient number of times produces the distributions of the mass ratios $M^{}_{2}/M^{}_{1}$ and $M^{}_{3}/M^{}_{1}$, which are shown in Figure~\ref{fig1} (a) and (b).
It can be seen that these ratios typically lie within a factor of a few, around 2 and 3, indicating that the right-handed neutrino masses tend not to exhibit a strong hierarchy. In this case, it is generally not appropriate to assume that leptogenesis is dominated solely by the decay of the lightest right-handed neutrino. Instead, the contributions from the two heavier right-handed neutrinos may play an important role in generating the baryon asymmetry. We therefore include the contributions from all three heavy neutrinos in our calculations.

We then generate the dimensionless components of $M^{}_{\rm D}$. Once the Dirac and Majorana mass matrices are specified up to the overall scales $D^{}_0$ and $M^{}_0$, the effective mass matrix of light neutrinos is obtained via the type-I seesaw relation given in Eq.~(1), with an overall scale factor of $D^2_0/M^{}_0$. We fix $D^2_0/M^{}_0$ by requiring that the resulting $M^{}_\nu$ reproduces $\Delta m^2_{21}$ in agreement with its measured value.

\begin{figure*}
\centering
\includegraphics[width=5.0in]{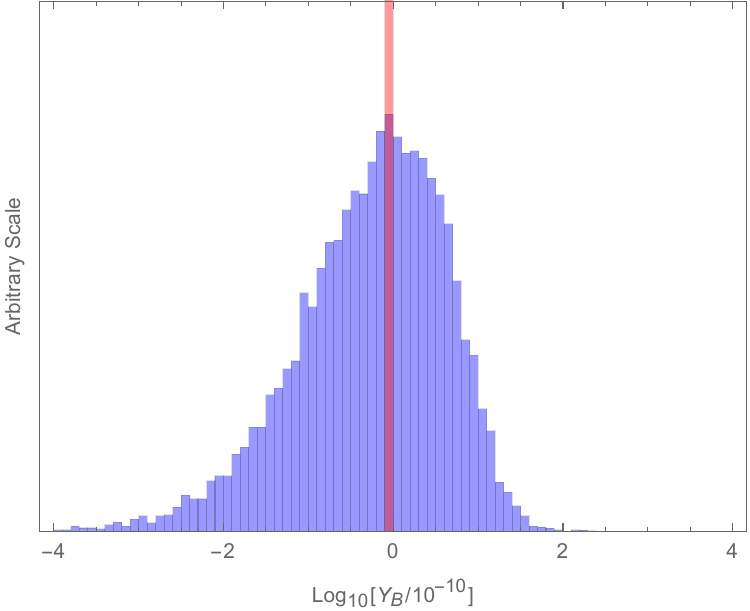}
\caption{For the scenario studied in section~2.1, the probability distribution of the baryon asymmetry $Y^{}_{\rm B}$ for $M^{}_0 = 5\times 10^{11}$ GeV.}
\label{fig2}
\end{figure*}

With $D^2_0/M^{}_0$ fixed, we are left with only one free parameter. We take $M^{}_0$ as this parameter and calculate the generated baryon asymmetry as a function of $M^{}_0$, including the contributions from all three heavy neutrinos and flavor effects. Figure~\ref{fig2} shows the probability distribution of $Y^{}_{\rm B}$ for $M^{}_0 = 5 \times 10^{11}$ GeV. Through scanning over different values of $M^{}_0$, we find that this particular choice leads the distribution to peak at the observed baryon asymmetry, which thus maximizes the likelihood for $Y^{}_{\rm B}$ to agree with the experimental value.
Notably, the probability distribution of $Y^{}_{\rm B}$ is rather concentrated, with the vast majority of results lying within one order of magnitude of the observed value.
This indicates that such a mass scale is favorable for successful leptogenesis within the random mass matrix framework. In comparison, the unflavored regime in Ref.~\cite{Lu:2014cla} typically prefers a somewhat higher scale, whereas our study demonstrates that successful leptogenesis can also be realized at relatively lower scales.

\subsection{Random generation of mass matrices with experimental cuts}

In this subsection, we impose experimental cuts on the randomly generated mass matrices, only retaining those combinations that reproduce neutrino mixing parameters consistent with the latest global fits. This ensures our analysis is constrained by real-world experimental measurements.

For this purpose, the Casas--Ibarra parametrization \cite{CI1, CI2} of $M^{}_{\rm D}$ provides a convenient means to incorporate the low-energy neutrino data. With the help of Eqs.~(\ref{1}, \ref{2}), one can easily understand that the Casas-Ibarra parametrization of $M^{}_{\rm D}$ takes a form as
\begin{eqnarray}
M^{}_{\rm D} = {\rm i} U D^{1/2}_\nu O D^{1/2}_{\rm R} \; ,
\label{2.7}
\end{eqnarray}
with $\sqrt{D^{}_\nu} = {\rm diag}(\sqrt{m^{}_1}, \sqrt{m^{}_2}, \sqrt{m^{}_3})$ and $\sqrt{D^{}_{\rm R}} = {\rm diag}(\sqrt{M^{}_1}, \sqrt{M^{}_2}, \sqrt{M^{}_3})$. Here $O$ contains the extra degrees of freedom of $M^{}_{\rm D}$ relative to $M^{}_\nu$ and satisfies the orthogonal relation $O^{T} O=I$.
The key advantage of this parametrization is that it separates the experimentally measured neutrino parameters from the unconstrained degrees of freedom.
Consequently, the former can be fixed to their observed values, while the latter are treated as free parameters in our numerical analysis.

For the orthogonal matrix $O$, a statistically consistent random generation procedure is required. As discussed in Ref.~\cite{DiBari:2018fvo}, a proper approach must account for the group structure of $O(3,\mathbb{C})$. Unlike unitary matrices, which belong to compact groups, the complex orthogonal group is non-compact, implying that its parameters are not confined to finite intervals. Consequently, a naive scan over complex angles does not yield a statistically well-defined random ensemble. A resolution is provided by the decomposition of a generic complex orthogonal matrix into a real rotation and a complex boost transformation in flavor space,
\begin{eqnarray}
O = R(\alpha^{}_{12},\alpha^{}_{13},\alpha^{}_{23})\, O^{}_{\rm boost}(\vec{\beta})\; ,
\label{2.8}
\end{eqnarray}
where $R$ is a real orthogonal matrix parametrized by three Euler angles, and $O^{}_{\rm boost}$ denotes a Lorentz-like boost characterized by a vector $\vec{\beta}$ in flavor space.
This decomposition offers a convenient parametrization of the most general complex orthogonal matrix. In our numerical analysis, the corresponding parameters are generated randomly such that the orthogonal matrices scan the full parameter space without introducing any preferred direction.

To obtain the right-handed neutrino masses, we randomly generate the Majorana mass matrix $M^{}_{\rm R}$ according to the Gaussian measure described in the previous subsection. The physical heavy neutrino masses $M^{}_1$, $M^{}_2$ and $M^{}_3$ are then obtained by diagonalizing $M^{}_{\rm R}$. In this way, the heavy neutrino mass spectrum emerges naturally from the random matrix structure, without assuming any particular hierarchy. This procedure allows the Casas--Ibarra parametrization to consistently combine the low-energy neutrino data with a fully random heavy neutrino sector, providing a statistically well-defined framework for studying leptogenesis.

\begin{figure*}
\centering
\includegraphics[width=6.5in]{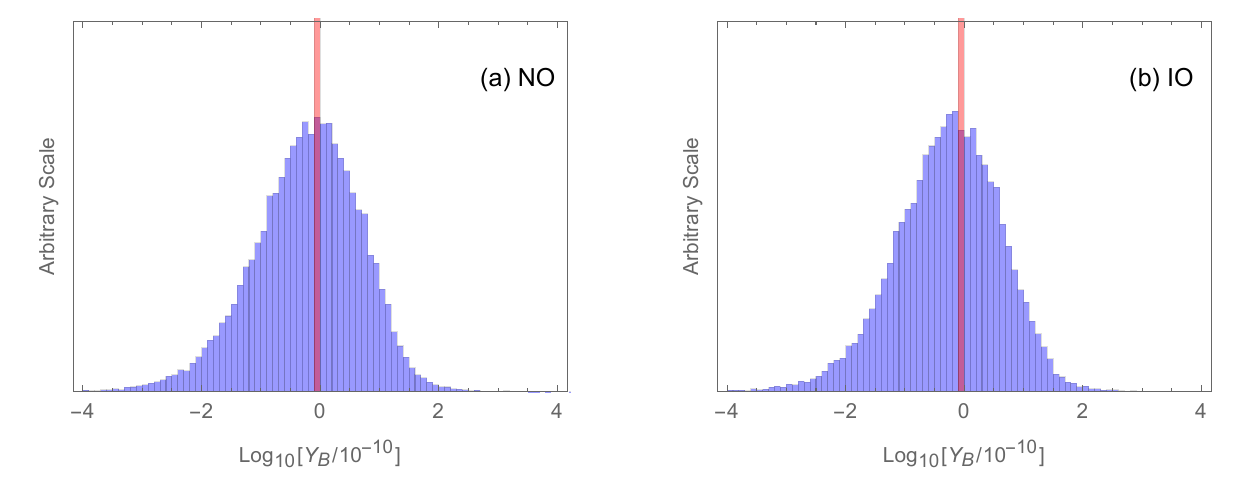}
\caption{For the scenario studied in section~2.2, the probability distributions of the baryon asymmetry $Y^{}_{\rm B}$ for $M^{}_0 = 7 \times 10^{11}$ GeV and $9 \times 10^{11}$ GeV in the NO (a) and IO (b) cases. }
\label{fig3}
\end{figure*}

Now, we are ready to perform the numerical calculations. For different values of the right-handed neutrino mass scale $M^{}_0$, we perform statistical scans over randomly generated mass matrices to obtain the probability distribution of the resulting baryon asymmetry $Y^{}_{\rm B}$. Figure~\ref{fig3} (a) and (b) (for the normal ordering (NO) and inverted ordering (IO) cases of light neutrino masses, respectively) show the probability distributions of $Y^{}_{\rm B}$ for $M^{}_0$ values searched and identified to make the peak of the $Y^{}_{\rm B}$ probability distribution consistent with the observed value.
These results are obtained for the following parameter settings: for the neutrino mass squared differences and neutrino mixing angles, we employ the global-fit results for them \cite{global1, global2}; for the Dirac CP phase and the two Majorana CP phases of the neutrino mixing matrix, we allow them to vary uniformly in the range 0---$2\pi$ (in light of that the neutrino mixing matrix follows the Haar measure \cite{Haba:2000be}); the lightest neutrino mass ($m^{}_1$ or $m^{}_3$ in the NO or IO case) is varied in the range between $0.001$ eV and $0.1$ eV; for the orthogonal matrix $O$ and the right-handed neutrino masses, we generate them in accordance with the discussions in the previous two sections.
Our results reveal that for the NO case, the probability of reproducing the observed value of $Y^{}_{\rm B}$ peaks at $M^{}_0 \sim 7 \times 10^{11}$ GeV, while for the IO case, the corresponding peak occurs at a slightly higher mass scale of $M^{}_0 \sim 9 \times 10^{11}$ GeV.

\section{Leptogenesis in the type-II seesaw}

In this section, we extend our analysis to the type-II seesaw model.

Unlike the type-I seesaw model, in which light neutrino masses arise from the exchange of heavy right-handed neutrinos, the type-II seesaw generates neutrino masses through the introduction of a massive scalar triplet $\Delta$ that couples directly to the lepton doublets and to the Higgs doublet \cite{Magg:1980ut}- \cite{Schechter:1980gr}. The relevant terms in the Lagrangian can be written as
\begin{eqnarray}
  \mathcal{L}_\Delta\ =\ -\,\frac{1}{2} \left( f_{\alpha\beta}\, L_\alpha^{T} C {\rm i} \sigma^2 \Delta L_\beta
    +\mu\, H^{T} {\rm i} \sigma^2 \Delta^\dagger H + \mbox{h.c.} \right) - M_\Delta^2\, \mbox{tr} (\Delta^\dagger \Delta) \; ,
\label{3.1}
\end{eqnarray}
where $f^{}_{\alpha\beta}$ is a symmetric matrix of complex couplings, and $\mu$ is a complex mass parameter. The exchange of the heavy scalar triplet generates a contribution to the light neutrino mass matrix, which can be written as
\begin{eqnarray}
  (m^{}_{\Delta})_{\alpha\beta}\ =\ \frac{1}{2}\, \mu f^{}_{\alpha\beta}\frac{v^2}{M^{2}_\Delta} \;.
\label{3.2}
\end{eqnarray}

In addition to generating neutrino masses, the scalar triplet can also produce a lepton asymmetry via its out-of-equilibrium decays into lepton pairs and Higgs pairs. However, within the minimal type-II seesaw framework, the CP asymmetry in the decays of the scalar triplet vanishes at the one-loop level~\cite{ODonnell:1993obr, Ma:1998dx}. Therefore, to generate a nonzero CP asymmetry, an additional heavy state (either another scalar triplet or a right-handed neutrino) is generally required, which also provides an additional contribution to the neutrino mass matrix
\begin{eqnarray}
  (m^{}_{\rm H})_{\alpha\beta}=\frac{1}{2}\kappa^{}_{\alpha\beta}\frac{v^2}{M^{}_{\rm H}} \; ,
\label{3.4}
\end{eqnarray}
with $M^{}_{\rm H}$ denoting its mass.

In the type-II leptogenesis scenario, the lepton asymmetry is generated through CP-violating decays of the scalar triplet, which at the tree level can decay into two leptons $\Delta \to {\overline L^{}_\alpha} + {\overline L^{}_\beta}$ or two SM Higgs doublets $\Delta \to H + H$. The resulting CP asymmetries are generally flavor dependent, arising from the interference of tree-level and one-loop diagrams. This leads to a nontrivial flavor structure in the generated lepton asymmetry, making the single-flavor approximation become insufficient. As a result, flavor effects play an essential role in the evolution of the lepton asymmetry and are still relevant even at temperatures where charged-lepton Yukawa interactions are out of equilibrium. To properly account for these effects, we adopt the flavor-covariant formalism of Ref.~\cite{Lavignac:2015gpa}. The corresponding flavor-covariant CP asymmetries are given by
\begin{eqnarray}
  \mathcal{E}_{\alpha\beta}\, =\, \frac{1}{8\pi {\rm i}}\frac{M_\Delta}{v^2}\sqrt{B_L B_H}\
    \frac{(m_H\, m_\Delta^{\dagger} - m_\Delta m_H^{\dagger})_{\alpha\beta}}{\sqrt{\text{tr}(m_\Delta^\dagger m_\Delta)}}\, .
\label{3.5}
\end{eqnarray}
Here, the branching ratios of the scalar triplet are defined as
\begin{eqnarray}
  B_L\, \equiv\, \mathrm{BR}(\Delta\rightarrow\bar L\bar L)\, =\, \frac{\lambda_L^2}{\lambda_L^2+\lambda_H^2}\ ,
  \qquad
  B_H\, \equiv\, \mathrm{BR}(\Delta\rightarrow HH)\, =\, \frac{\lambda_H^2}{\lambda_L^2+\lambda_H^2}\ ,
\label{3.6}
\end{eqnarray}
with $\lambda_L\, \equiv\, \sqrt{\text{tr}(ff^\dagger)}$ and $\lambda_H\, \equiv\, \frac{|\mu|}{M_\Delta}$.

In this work, we focus on the temperature range $10^9 - 10^{12}$ GeV, where the $\tau$-Yukawa interactions are in equilibrium while the $e$ and $\mu$ flavors remain coherent, corresponding to the two-flavor regime. In this regime, the $\tau$ Yukawa coupling is in equilibrium and drives the $(e,\tau)$, $(\mu,\tau)$, $(\tau,e)$ and $(\tau,\mu)$ entries of the $3\times 3$ density matrix $(\Delta_L)_{\alpha\beta}$ vanish. As a result, the flavor density matrix effectively decomposes into a $2\times2$ submatrix in the $(e,\mu)$ sector and a separate $\tau$ flavor. Accordingly, the lepton asymmetry matrix $\Delta_{\alpha\beta}$ is described by $\Delta^{0}_{\alpha\beta}$ and $\Delta_{\tau}$. The corresponding relations among the asymmetry densities are given by
\begin{eqnarray}
&&  (\Delta^{0}_{L})_{\alpha\beta}=\left(\frac{86}{589}{\rm Tr}(\Delta^{0}_{\alpha\beta})
 +\frac{60}{589}\Delta_{\tau}+\frac{8}{589}\Delta_{\Delta} \right)\delta_{\alpha\beta}-\Delta^{0}_{\alpha\beta},
  \nonumber \\
&&  \Delta_{L_{\tau}}=\frac{30}{589} {\rm Tr}(\Delta^{0}_{\alpha\beta})-\frac{390}{589}\Delta_{\tau}
 -\frac{52}{589}\Delta_{\Delta},
 \nonumber \\
&&  \Delta_{\phi}=-\frac{164}{589}{\rm Tr}\left(\Delta^{0}_{\alpha\beta}\right)
 -\frac{224}{589}\Delta_{\tau}-\frac{344}{589}\Delta_{\Delta}.
 \label{3.7}
\end{eqnarray}
The dynamical variables of scalar triplet leptogenesis are $\Sigma_\Delta \equiv (n_\Delta+n_{\bar \Delta})/s$, $\Delta_\Delta \equiv (n_\Delta-n_{\bar \Delta})/s$ and the lepton asymmetry matrix $\Delta_{\alpha\beta}$. The evolution of these quantities is governed by the flavor-covariant Boltzmann equations
\begin{eqnarray}
&&  sHz\frac{d\Sigma_\Delta}{dz}\,  =\, -\left(\frac{\Sigma_\Delta}{\Sigma_\Delta^\text{eq}}-1\right)\! \gamma_D
    -2\left(\! \left( \frac{\Sigma_\Delta}{\Sigma_\Delta^\text{eq}} \right)^{\! 2} - 1\right)\! \gamma_A \; ,
 \nonumber \\
&&  sHz\frac{d\Delta^0_{\alpha\beta}}{dz}\,  =\,
    -\left(\frac{\Sigma_\Delta}{\Sigma_\Delta^\text{eq}}-1\right)\! \gamma_D\, \mathcal{E}_{\alpha\beta}
    +\tilde{\mathcal{W}}^D_{\alpha\beta}+\tilde{\mathcal{W}}^{L H}_{\alpha\beta}+\tilde{\mathcal{W}}^{4L}_{\alpha\beta}
    +\tilde{\mathcal{W}}^{L\Delta}_{\alpha\beta} \; ,
 \nonumber \\
&&  sHz\frac{d\Delta_\tau}{dz}\,  =\,
    -\left(\frac{\Sigma_\Delta}{\Sigma_\Delta^\text{eq}}-1\right)\! \gamma_D\, \mathcal{E}_{\tau\tau}
    +\tilde{\mathcal{W}}^D_{\tau}+\tilde{\mathcal{W}}^{L H}_{\tau}+\tilde{\mathcal{W}}^{4L}_{\tau}
    +\tilde{\mathcal{W}}^{L\Delta}_{\tau} ,
 \nonumber \\
&&  sHz\frac{d\Delta_\Delta}{dz}\,  =\, -\frac{1}{2}\left(\text{tr}(\tilde{\mathcal{W}}^D)+\tilde{\mathcal{W}}^D_\tau-W^D_H\right) \;,
\label{3.8}
\end{eqnarray}
with $z\equiv M_{\Delta}/T$, $H =1.66 \sqrt{g_*}\, T^2/M^{}_{\rm Pl}$ the Hubble rate, $g^{}_*$ the number of relativistic degrees of freedom and $M^{}_{\rm Pl} = 1.22\times 10^{19}$ GeV the Planck mass. The explicit expressions of the washout terms $\tilde{\mathcal{W}}^{\, \cdots}_{\alpha\beta}$,
$\tilde{\mathcal{W}}^{\, \cdots}_{\tau}$ and $W^D_H$ are given in Appendix.

We now investigate leptogenesis in the type-II seesaw model within the framework of neutrino mass anarchy, with flavor effects consistently taken into account. Recall that the neutrino mass contributions from the scalar triplet and the additional heavy state are described by the mass matrices $m^{}_{\Delta}$ and $m^{}_{\rm H}$, respectively, whose sum yields the full light neutrino mass matrix $M^{}_\nu$. In analogy to the treatment of $M^{}_{\rm D}$ and $M^{}_{\rm R}$ in section~2.1, $m^{}_{\Delta}$ ($m^{}_{\rm H}$) can be parameterized as the product of an overall mass scale $m^{0}_{\Delta}$ ($m^{0}_{H}$) and a matrix consisting of dimensionless random elements that follow the Gaussian measure.

\begin{figure*}
\centering
\includegraphics[width=6.5in]{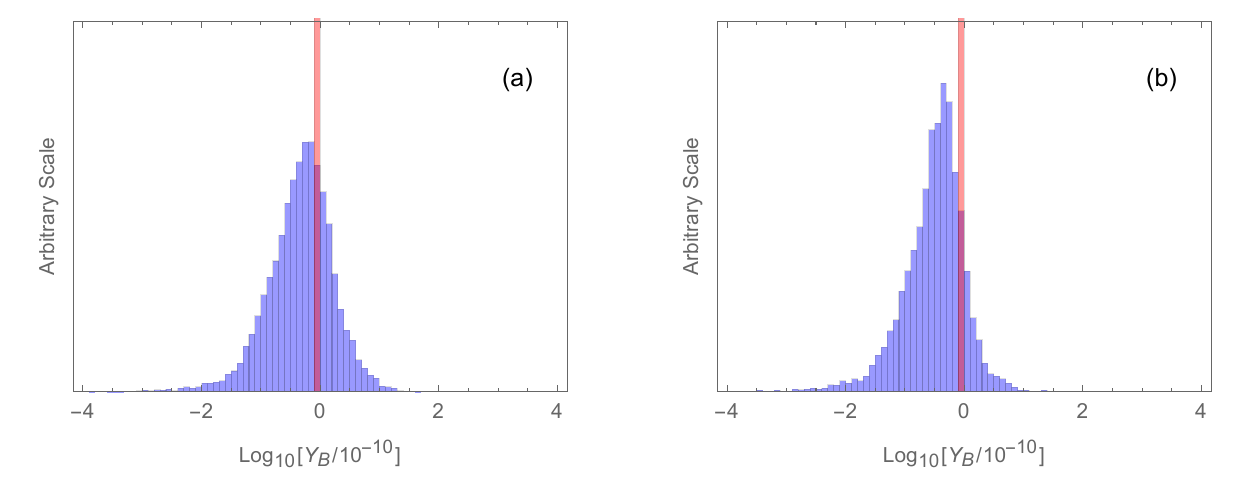}
\caption{ For the scenario studied in section~3, the probability distributions of the baryon asymmetry $Y^{}_{\rm B}$ for $M^{}_\Delta = 3\times 10^{11}$ GeV in the cases of $M^{}_{\Delta} = M^{}_{\rm H}/3$ (a) and $M^{}_{\Delta} = M^{}_{\rm H}$ (b).}
\label{fig4}
\end{figure*}

Given that the contribution of a heavy state to the light neutrino masses in the seesaw mechanism is generically inversely proportional to its mass scale, it is natural to adopt the relation $m^{0}_{\Delta}/m^{0}_{\rm H} \sim M^{}_{\rm H}/M^{}_\Delta$.
Therefore, varying the ratio $M^{}_{\Delta}/M^{}_{\rm H}$ effectively controls the relative importance of the triplet-induced and heavy-state-induced contributions to the light neutrino mass matrix. For different values of the triplet mass scale $M^{}_{\Delta}$, we perform statistical scans by randomly generating the corresponding mass matrices and calculate the resulting baryon asymmetry $Y^{}_{\rm B}$. Figure~\ref{fig4} (a) and (b) display the probability distributions of $Y^{}_{\rm B}$ for two typical choices of the ratio between the triplet and heavy mass scales, namely $M^{}_{\Delta} = M^{}_{\rm H}/3$ and $M^{}_{\Delta} = M^{}_{\rm H}$, respectively.
Our results show that, in both cases, the observed value of $Y^{}_{\rm B}$ can be most naturally reproduced for $M^{}_{\Delta}$ at the scale of $O(10^{11})$ GeV, particularly around $M^{}_{\Delta} \sim 3 \times 10^{11}$ GeV. A slight shift of the peak toward lower values of $Y^{}_{\rm B}$ is observed in Figure~\ref{fig4} (b) relative to Figure~\ref{fig4} (a). This feature can be attributed to the suppressed relative contribution of the scalar triplet at larger values of the ratio $M^{}_{\Delta}/M^{}_{\rm H}$.

\section{Summary}

Understanding the origin of neutrino masses and the baryon-antibaryon asymmetry of the Universe represents one of the most fundamental open issues in particle physics and cosmology. Neutrino oscillation experiments have firmly established that neutrinos are massive and lepton flavors are mixed, while the seesaw mechanism provides an elegant theoretical framework to account for the smallness of neutrino masses and simultaneously realize leptogenesis as the origin of the observed baryon asymmetry. In this context, the neutrino mass anarchy hypothesis offers a general and model-independent approach, where neutrino mass matrices are treated as random structureless matrices without imposing any ad hoc flavor symmetries.

In this work, we have performed a systematic investigation of leptogenesis within the neutrino mass anarchy framework, covering both the type-I and type-II seesaw scenarios with flavor effects consistently incorporated throughout the analysis. For the type-I seesaw model, two complementary setups have been studied: a fully anarchic scenario where both the Dirac and Majorana mass matrices are generated randomly without experimental constraints, and a phenomenologically constrained scenario based on the Casas-Ibarra parametrization that reproduces the measured neutrino oscillation observables. In both setups, contributions from all three right-handed neutrinos are included, instead of assuming dominance by the lightest one. For the type-II seesaw model, we have extended the anarchy hypothesis to the scalar triplet and additional heavy state sectors, where leptogenesis is driven by the out-of-equilibrium and CP-violating decays of the scalar triplet.

For the type-I seesaw scenario, the optimal mass scales that maximize the probability of matching the observed baryon asymmetry lie in the two-flavor regime, ranging from $10^{11}\,{\rm GeV}$ to $10^{12}\,{\rm GeV}$. Specifically, the fully random setup favors $M^{}_{0}\sim 5\times 10^{11}\,{\rm GeV}$, while the Casas-Ibarra constrained setup yields peak probabilities at $M^{}_{0}\sim 7\times 10^{11}\,{\rm GeV}$ for normal ordering and $M^{}_{0}\sim 9\times 10^{11}\,{\rm GeV}$ for inverted ordering of light neutrino masses. For the type-II seesaw scenario, the preferred mass scale of the scalar triplet is slightly lower, around $M^{}_{\Delta}\sim 3\times 10^{11}\,{\rm GeV}$, and the relative contribution between the triplet and the additional heavy state can affect the shape of the $Y^{}_{\rm B}$ distribution.

\vspace{0.5cm}

\begin{appendix}

\section*{Appendix}

The space-time density of total decay of the triplet scalar $\Delta$ and its anti-particle $\overline{\Delta}$ is given by
\begin{equation}
\gamma_D=s\Gamma_\Delta\Sigma^{eq}_\Delta\frac{K_1(z)}{K_2(z)} \;,
\end{equation}
where $\Gamma_\Delta\, =\, \frac{1}{32\pi} \left( \lambda^2_L + \lambda^2_H \right) M_\Delta $ is the triplet decay width, $K_{1, 2} (z)$ are the modified Bessel functions of the second kind. For generic $2 \to 2$ scatterings, the corresponding reaction density is given by
\begin{equation}
  \gamma(a+b\rightarrow i+j)\, =\, \frac{T}{64\pi^4}\int_{s_\mathrm{min}}^\infty\! ds\, s^{1/2}\hat{\sigma}(s)\,
    K_1\! \left(\frac{\sqrt{s}}{T}\right) .
\end{equation}
In particular, the gauge-induced scattering contribution reads
\begin{equation}
\gamma_A=\frac{z T^4e^{-2z}}{64\pi^4}(9g^4
 +12g_2^2g^2_{\mathcal{Y}}+3g^4_Y)\left(1+\frac{3}{4z}\right),
 \label{eq:gammaA}
\end{equation}
with $g_2 =\frac{e}{\sin\theta_W} \simeq 0.65$ and $g_{\mathcal{Y}}=\frac{e}{\cos\theta_W} \simeq 0.46$.

The washout term $W^D_{\alpha \beta}$ associated with triplet and anti-triplet inverse decays is given by
\begin{align}
  \mathcal{W}^D_{\alpha\beta}\, =\, \frac{2B_L}{\lambda_L^2}
    \left[ (ff^\dagger)_{\alpha\beta}\frac{\Delta_\Delta}{\Sigma_\Delta^{\text{eq}}} + \frac{1}{4Y_L^{\text{eq}}}
    \left( 2f\Delta_L^Tf^\dagger + ff^\dagger\Delta_L + \Delta_L ff^\dagger \right)_{\alpha\beta}\right]\! \gamma_D \, .
\label{eq:W_D}
\end{align}
In addition to decays and inverse decays, several $2 \to 2$ scattering processes contribute to the evolution of the lepton asymmetry. The contribution from $\Delta L =2$ scatterings $L_\gamma L_\delta \leftrightarrow \bar H \bar H$
and $L_\gamma H \leftrightarrow \bar L_\delta \bar H$ are described by
\begin{align}
 \mathcal{W}^{ L H}_{\alpha\beta}\, =\ & 2\left\lbrace\frac{1}{\lambda^2_L}\left[\frac{\left(2f\Delta_L^Tf^\dagger+ff^\dagger\Delta_L+\Delta_L ff^\dagger\right)_{\alpha\beta}}{4Y_L^\text{eq}}+\frac{\Delta_H}{Y_H^{\text{eq}}}(ff^\dagger)_{\alpha\beta}\right]\right.\! \gamma_{L H}^\Delta\nonumber\\
&+\frac{1}{\mbox{Re}\left[\text{tr}(f\kappa^\dagger)\right]}\left[\frac{\left(2f\Delta_L^T\kappa^\dagger+f\kappa^\dagger\Delta_L+\Delta_L f\kappa^\dagger\right)_{\alpha\beta}}{4Y_L^\text{eq}}+\frac{\Delta_H}{Y_H^{\text{eq}}}(f\kappa^\dagger)_{\alpha\beta}\right]\! \gamma_{L H}^\mathcal{I}\nonumber\\
&+\frac{1}{\mbox{Re}\left[\text{tr}(f\kappa^\dagger)\right]}\left[\frac{\left(2\kappa\Delta_L^T f^\dagger+\kappa f^\dagger\Delta_L+\Delta_L\kappa f^\dagger\right)_{\alpha\beta}}{4Y_L^\text{eq}}+\frac{\Delta_H}{Y_H^{\text{eq}}}(\kappa f^\dagger)_{\alpha\beta}\right]\! \gamma_{L H}^\mathcal{I}\nonumber\\
&\left.+\frac{1}{\lambda^2_\kappa}\left[\frac{\left(2\kappa\Delta_L^T\kappa^\dagger+\kappa\kappa^\dagger\Delta_L+\Delta_L\kappa\kappa^\dagger\right)_{\alpha\beta}}{4Y_L^\text{eq}}+\frac{\Delta_H}{Y_H^{\text{eq}}}(\kappa\kappa^\dagger)_{\alpha\beta}\right]\! \gamma_{L H}^\mathcal{H}\right\rbrace,
\label{eq:W_lh}
\end{align}
where $\Delta_H \equiv (n_H-n_{\bar H})/s$,
$\lambda_\kappa \equiv \sqrt{\text{tr}(\kappa\kappa^\dagger)}$
, $\gamma_{L H}^\Delta$, $\gamma_{L H}^\mathcal{H}$ and $\gamma_{L H}^\mathcal{I}$
are the contributions of different self-energy diagrams, and the total reaction density satisfies
$\gamma_{L H}=\gamma_{L H}^\Delta+2\gamma_{L H}^\mathcal{I}+\gamma_{L H}^\mathcal{H}$.
The washout term due to the 2 lepton--2 lepton scatterings $L_\alpha L_\beta\leftrightarrow L_\gamma L_\delta$
and $L_\alpha\bar L_\gamma\leftrightarrow\bar L_\beta L_\delta$ is given by
\begin{align}
\mathcal{W}^{4L}_{\alpha\beta}\, =\, \frac{2}{\lambda^4_L}&\left[ \lambda^2_L\, \frac{\left(2f\Delta_L^Tf^\dagger+ff^\dagger\Delta_L+\Delta_L ff^\dagger\right)_{\alpha\beta}}{4Y_L^\text{eq}}
- \frac{\text{tr}(\Delta_L ff^\dagger)}{Y_L^{\text{eq}}}(ff^\dagger)_{\alpha\beta}\right]\! \gamma_{4L} \,.
\label{eq:W_4l}
\end{align}
The contribution from the lepton-triplet scatterings $L_\gamma\Delta\leftrightarrow L_\delta\Delta$,
$L_\gamma\bar \Delta\leftrightarrow L_\delta\bar \Delta$ and
$L_\gamma\bar L_\delta\leftrightarrow\Delta\bar \Delta$ is
\begin{align}
\mathcal{W}^{L\Delta}_{\alpha\beta}\, =\, \frac{1}{\text{tr}(ff^\dagger ff^\dagger)}\left[\frac{1}{2Y_L^\text{eq}}\left(ff^\dagger ff^\dagger\Delta_L-2ff^\dagger\Delta_L ff^\dagger+\Delta_L ff^\dagger ff^\dagger\right)_{\alpha\beta}\right]\! \gamma_{L\Delta}\, .
\label{eq:W_ellDelta}
\end{align}
Finally, the washout term associated with Higgs inverse decays $H H \to \Delta$ and $\bar H \bar H \to \bar \Delta$ is given by
\begin{align}
  W^D_H\, =\, 2B_H\left(\frac{\Delta_H}{Y_H^\text{eq}}-\frac{\Delta_\Delta}{\Sigma_\Delta^\text{eq}}\right)\! \gamma_D\, .
\label{eq:W^D_H}
\end{align}

The corresponding washout terms $\tilde{\mathcal{W}}^{\, \cdots}_{\alpha\beta}$
$(\alpha, \beta = e, \mu)$ and $\tilde{\mathcal{W}}^{\, \cdots}_{\tau}$ appearing in Eq.~(\ref{3.8}) are obtained by setting
$(\Delta_L)_{\alpha\tau} = (\Delta^*_L)_{\tau\alpha} = \Delta_{L_\tau} \delta_{\alpha\tau}$
in the expressions
of the corresponding $\mathcal{W}^{\, \cdots}_{\alpha\beta}$ $(\alpha, \beta = e, \mu, \tau)$. The explicit expressions of the reduced cross section $\hat{\sigma}(s)$ for the scattering processes discussed above can be found in Ref.~\cite{Lavignac:2015gpa}.

\end{appendix}

\vspace{0.5cm}

\underline{Acknowledgments} \vspace{0.2cm}

This work was supported in part by the National Natural Science Foundation of China under Grant No. 12475112, Liaoning Revitalization Talents Program under Grant No. XLYC2403152, and the Basic Research Business Fees for Universities in Liaoning Province under Grant No. LJ212410165050.

\end{document}